\documentclass[preprint]{aastex631}
\usepackage{amsmath}

\shorttitle{Reconnection Acceleration with Feedback}
\shortauthors{Seo et al.}

\begin{document}

\title{Proton Acceleration in Low-$\beta$ Magnetic Reconnection with Energetic Particle Feedback}
\author[0000-0002-5550-8667]{Jeongbhin Seo}
\affiliation{Los Alamos National Laboratory, Los Alamos, NM 87545, USA}
\author[0000-0003-4315-3755]{Fan Guo}
\affiliation{Los Alamos National Laboratory, Los Alamos, NM 87545, USA}
\affiliation{New Mexico Consortium, Los Alamos, NM 87544, USA}
\author[0000-0001-5278-8029]{Xiaocan Li}
\affiliation{Dartmouth College, Hanover, NH 03750 USA}
\author[0000-0003-3556-6568]{Hui Li}
\affiliation{Los Alamos National Laboratory, Los Alamos, NM 87545, USA}

\begin{abstract}
Magnetic reconnection regions in space and astrophysics are known as active particle acceleration sites. There is ample evidence showing that energetic particles can take a substantial amount of converted energy during magnetic reconnection. However, there has been a lack of studies understanding the backreaction of energetic particles at magnetohydrodynamical scales in magnetic reconnection. To address this, we have developed a new computational method to explore the feedback by non-thermal energetic particles. This approach considers the backreaction from these energetic particles by incorporating their pressure into Magnetohydrodynamics (MHD) equations. The pressure of the energetic particles is evaluated from their distribution evolved through Parker's transport equation, solved using stochastic differential equations (SDE), so we coin the name MHD-SDE. Applying this method to low-$\beta$ magnetic reconnection simulations, we find that reconnection is capable of accelerating a large fraction of energetic particles that contain a substantial amount of energy. When the feedback from these particles is included, their pressure suppresses the compression structures generated by magnetic reconnection, thereby mediating particle energization. Consequently, the feedback from energetic particles results in a steeper power-law energy spectrum. These findings suggest that feedback from non-thermal energetic particles plays a crucial role in magnetic reconnection and particle acceleration.

\end{abstract}

\section{Introduction}
Magnetic reconnection plays a crucial role in magnetic energy release and particle acceleration in various astrophysical and heliophysical phenomena \citep{zweibel2009,yamada2010,guo2014,li2021,ji2022,guo2024}, such as solar flares \citep{li2022,shen2022}, astrophysical jets \citep{romanova1992,giannios2009}, magnetotails \citep{nagai2001,angelopoulos2008,oka2023}, and the heliospheric current sheet \citep{desai2022,zhang2024,murtas2024}. In many of these phenomena, energetic particles play a significant role. For example, in the looptop regions of solar flares \citep{kruker2010, oka2013, oka2015}, non-thermal electrons driven by magnetic reconnection can account for a substantial fraction of the total electron energy, roughly $\sim50\%$. While the feedback of cosmic rays has been extensively studied in supernova shocks \citep[e.g.,][]{drury1981,axford1982,chevalier1983, bell2013, cristofari2021}, there has been less focus on magnetic reconnection \citep{drake2019}.

Simulation studies of particle acceleration utilize various numerical methods, with one of the most well-known being the particle-in-cell (PIC) method. This method, based on the first principles Newton-Maxwell description \citep{dawson1983}, evolves Lagrangian macro-particles using fields that are solved via the Eulerian method. The PIC method is very useful for understanding particle acceleration in various astrophysical phenomena, including diffusive shock acceleration \citep{guo2014a,guo2014b,ha2021} and acceleration in the magnetic reconnection region \citep{drake2006,oka2010,Li2017,Li2019,Zhang2021,guo2015,guo2021}. However, the PIC method has to resolve the electron kinetic scales, making it prohibitively challenging for large-scale problems (e.g., solar flare regions and protostellar disks are larger than $10^9$ times of electron kinetic scales). To overcome this limitation, the hybrid-PIC method has been developed and utilized \citep{giacalone1992,Guo2013,Caprioli2013,Haggerty2020,winske2023,le2023,zhang2024}. This approach treats electrons as a fluid, and solves the kinetic motion of ions, so it only needs to resolve ion kinetic scales. Although this method extends the dynamical scale of kinetic simulations, its domain size is still limited to several thousand ion inertial lengths. 

Various methods have been developed for studying particle acceleration at even larger scales. Among these, the Monte-Carlo method is widely employed for studying particle acceleration on a hydrodynamic scale, as it directly tracks the transport of macro-particles in the background fluid with a particle scattering model. This method has been used to study particle acceleration in various astrophysical phenomena, including collisionless shock \citep{ellison1984, ellison1990, ellison1995, ellison2013}, turbulence \citep{ohira2013}, and shear flows \citep{ostrowski1998}. For example, \citet{seo2023b} utilized this method to simulate particle acceleration in radio galaxy jets, discussing the acceleration processes in complex flows that include shocks, shear, and turbulence. However, these studies typically do not include feedback from energetic particles.

The energetic particle transport theory is another approach for studying particle acceleration and transport on large scales \citep{parker1965,blanford1987,zank2014}. Various transport equations have been derived to study the dynamical evolution of energetic particle distributions \citep{parker1965,leroux2009,zank2014}. Traditionally, these equations have been solved using the finite difference method \citep{jokipii1979, potgieter1985, burger1995}. More recently, the stochastic differential equation (SDE) method has been widely adopted for solving the transport equation \citep{zhang1999, florinski2009, pei2010, kong2017, li2018}. This method has been applied to understand the transport and acceleration of energetic particles in the outer heliosphere \citep{florinski2009}, at coronal shocks \citep{kong2017}, in low-$\beta$ magnetic reconnection regions \citep{li2018}, and in solar flare regions \citep{kong2019, li2022}. While this approach is robust and allows for anisotropic spatial diffusion \citep{giacalone1999}, previous numerical studies have rarely considered the feedback of non-thermal energetic particles on the MHD fluid.

There are several approaches to taking into account the feedback of non-thermal fluid at the hydrodynamic scale \citep[see][for a review]{ruszkowski2023}. A common approach is to include the pressure of energetic particles into MHD equations \citep{kang2007,kudoh2016}. Recently, several new models have been developed to include particle dynamics and feedback. The $\it{kglobal}$ model \citep{drake2019,arnold2021} applies the momentum equation of the guiding center of energetic particles to the MHD equations. It is capable of addressing the effects of the instabilities (e.g., firehose instability) driven by the anisotropic pressure of energetic particles. However, this approach currently does not explicitly include the scattering of energetic particles by magnetic turbulence. The MHD-PIC method \citep{bai2015} includes the backreaction of the Lorentz force from Lagrangian cosmic ray particles in the fluid momentum and energy equations. This approach needs to resolve the gyromotion scale of energetic particles.
\citet{wang2022} solved the coupled MHD equation, turbulence transport equation, and cosmic rays fluid equation to investigate the backreaction. In their work, the presence of energetic particles introduces not only an isotropic pressure but also a viscous stress tensor into the dynamical equations. Besides, the self-generated turbulence by energetic particles due to the streaming instability also introduces dynamical feedback.

In this study, we introduce a new method called MHD-SDE, which incorporates the feedback of non-thermal energetic particles into the MHD framework by solving Parker's transport equation using SDE. We apply this method to proton acceleration in magnetic reconnection within solar flare environments with a weak guide field, where this approach is suitable.
The Parker's transport equation assumes frequent pitch-angle scattering that maintains a nearly isotropic momentum distribution, and includes the spatial diffusion and compression acceleration of energetic particles. By first principle PIC simulations, \cite{li2018a} showed that compression energization is crucial for accelerating high-energy particles during reconnection with a weak guide field. The particle distributions are nearly isotropic in the low guide field regime, and the acceleration due to flow shear is inefficient because it is proportional to the anisotropy level. Additionally, \cite{li2019a} showed that turbulence generated by 3D reconnection can efficiently scatter high-energy particle particles, leading to nearly isotropic particle distributions. Recent MHD simulations suggest that the reconnection layer is compressible, especially when plasma $\beta$ is low and the guide field is weak \citep{birn2012,provornikova2016}. \citet{dahlin2022} investigated the evolution of the guide field during an eruptive flare using 3D MHD simulations. They found that, even when the guide field starts strong, it weakens during the reconnection process and becomes nearly zero when the reconnection rate peaks. Additionally, \citet{chen2020} show that the weak guide field model better reproduces the magnetic field structure observed in the reconnection layer of the 2017 September 10th solar flare. Therefore, our simulations are highly relevant to solar flare observations.

We find that reconnection can accelerate a significant fraction of particles, and increased feedback from energetic particles results in weaker fluid compression in the reconnection layer, leading to a steeper power-law energy spectrum. We conduct a series of test problems using the MHD-SDE method to ensure its proper functionality. The method satisfies the Riemann solution, enabling its application on a large dynamic scale with non-thermal particle feedback.

The paper is organized as follows: Section \ref{s2} introduces the numerical method, followed by a discussion of the simulation results in Section \ref{s3}. A brief summary and an outlook are provided in Section \ref{s4}. In the Appendix, we included the results for the MHD-SDE method for a series of test problems.

\section{Numerical Method}
\label{s2}

Past numerous studies have incorporated the feedback of non-thermal fluids in MHD equations by including their pressure, a method akin to a ``two-fluid" approach \citep{drury1981,axford1982,kang2007,kudoh2016,ruszkowski2023,Habegger2024}. Recently, \citet{bai2015} and \citet{drake2019} suggested new approaches that add the force and work done by non-thermal particles directly into the MHD equations.
In Section \ref{s.2.1}, we will demonstrate how the MHD equations including feedback from non-thermal particles are derived under our assumptions, starting with the force density of the non-thermal fluid similar to \citet{drake2019}. We treat non-thermal particles as an isotropic distribution around the local fluid velocity in momentum space, as we use Parker's transport equation to solve for the transport and acceleration of non-thermal particles. Note that since we solve the pressure from the non-thermal fluid with a particle-based method and discuss non-thermal particle acceleration, we use ``non-thermal fluid'' and ``non-thermal particles'' interchangeably depending on the context.

We apply the MHD-SDE method to nonthermal proton acceleration in a solar flare environment.
In Section \ref{s.2.2.1}, we outline the setup and parameters for the MHD simulations of magnetic reconnection in a periodic box. Section \ref{s.2.2.2} covers the setup and parameters for the SDE part, including the initial conditions for non-thermal particles and the model of the spatial diffusion coefficient.

The MHD-SDE method provides remarkable features. In Appendix, we provide results for several test problems. This method accurately reproduces the Riemann solution in hydrodynamics (see Appendix \ref{A2} and \ref{A4}) and MHD (Appendix \ref{A3}), including non-thermal plasma pressure. Thanks to the high-order accuracy of the MHD code, discontinuities, including contact discontinuities, are resolved within 3-4 cells. By solving Parker's transport equation, we obtain the energy spectrum of the non-thermal fluid, successfully describing the analytical solution of first-order Fermi acceleration (see Section \ref{A5}). Therefore this method has the capability to investigate particle acceleration and feedback for different astrophysical and space physics problems.

\subsection{MHD-SDE Method}
\label{s.2.1}

The resistive MHD momentum and energy conservation equations, including force, $\mathbf{F_{\rm{NT}}}$, and work, $\mathbf{v}\cdot\mathbf{F_{\rm{NT}}}$, from the non-thermal fluid, can be written as \citep[e.g.,][]{bai2015},
\begin{eqnarray}
    \frac{\partial\rho_g\mathbf{v}}{\partial t} + \nabla\cdot\left(\rho_g \mathbf{v}\mathbf{v} + P^*\mathbf{I}-\mathbf{B}\mathbf{B}\right)  = \mathbf{F_{\rm{NT}}}, \\
    \frac{\partial E}{\partial t} + \nabla\cdot\left[(E+P^*)\mathbf{v}-\mathbf{B}(\mathbf{B}\cdot \mathbf{v})\right] = \nabla\cdot(\mathbf{B}\times\eta\mathbf{j})+\mathbf{v}\cdot\mathbf{F_{\rm{NT}}},
\end{eqnarray}
where
\begin{eqnarray}
    E=P_g/(\gamma_g-1) + (\rho_g \mathbf{v}\cdot\mathbf{v} + \mathbf{B}\cdot\mathbf{B})/2,
\end{eqnarray}
where $\rho_g$, $\mathbf{v}$, $\mathbf{B}$, $\gamma_g$, $\eta$, and $\mathbf{j}$ represent density, velocity, magnetic field, the adiabatic index of the thermal plasma, resistivity, and current density respectively. 
$\mathbf{B}$ is normalized by $\sqrt{4\pi}$ for convenience. $P^*$ is the sum of plasma gas pressure and magnetic pressure, $P^* = P_g + B^2/2$. In this context, the subscripts $g$ and ${\rm{NT}}$ refer to the thermal plasma gas and non-thermal fluid, respectively.
We consider the non-thermal component as a ``non-thermal fluid'', so the force density is given as \citep{drake2019},
\begin{equation}
    \mathbf{F_{\rm{NT}}} = - n_{\rm{NT}}\mathcal{E} - n_{\rm{NT}} \mathbf{u_{\rm{NT}}}\times\mathbf{B}/c - \nabla \cdot P_{\rm{NT}}\mathbf{I}, \label{Fcr}
\end{equation}
where $\mathbf{u_{\rm{NT}}}$ is the bulk velocity of the non-thermal fluid, $P_{\rm{\rm{NT}}}$ is the pressure of the non-thermal fluid, which can be calculated from the momentum distribution of the non-thermal particles (see Equations (\ref{pnt}) and (\ref{ent})). The electric field, $\mathcal{E}$, is given as MHD Ohm's law term and Hall term including non-thermal fluid \citep{bai2015},
\begin{equation}
    \mathcal{E} = -\frac{\mathbf{v}}{c}\times\mathbf{B} - \frac{n_{\rm{\rm{NT}}}}{|n_e|}\frac{(\mathbf{u_{\rm{NT}}} - \mathbf{v})}{c}\times\mathbf{B},
\end{equation}
where $n_e$ are the charge densities of electrons. In this simulation, we assume that the turbulence has no relative speed compared to the bulk flow, and non-thermal particles are assumed to have isotropic distribution around the local fluid velocity in the momentum space, so the velocity of the bulk motion of energetic particles is nearly the same as the thermal plasma velocity,
$\mathbf{u_{\rm{NT}}} \approx \mathbf{v}$, then Equation (\ref{Fcr}) becomes,
\begin{equation}
    \mathbf{F_{\rm{NT}}} = -\nabla \cdot P_{\rm{NT}}\mathbf{I}.
\end{equation}

Therefore, resistive MHD equations with non-thermal fluid feedback can be written as,

\begin{eqnarray}
    \frac{\partial \rho_g}{\partial t} + \nabla\cdot(\rho_g \mathbf{v}) = 0\label{cont},\\
    \frac{\partial\rho_g\mathbf{v}}{\partial t} + \nabla\cdot\left(\rho_g \mathbf{v}\mathbf{v} + P^*\mathbf{I}-\mathbf{B}\mathbf{B}\right) = -\nabla \cdot P_{\rm{NT}}\mathbf{I} \label{meq},\\
    \frac{\partial E}{\partial t} + \nabla\cdot\left[(E+P^*)\mathbf{v}-\mathbf{B}(\mathbf{B}\cdot \mathbf{v})\right] = \nabla\cdot(\mathbf{B}\times\eta\mathbf{j})-\mathbf{v}\cdot(\nabla \cdot P_{\rm{NT}}\mathbf{I})\label{eeq},\\
    \frac{\partial \mathbf{B}}{\partial t} - \nabla\times(\mathbf{v}\times\mathbf{B}) = \eta\nabla^2\mathbf{B}. \label{indu}
\end{eqnarray}
Equation (\ref{meq}) is identical to Equation (6) in \citet{drake2019} assuming the pressure is isotropic. The schemes for solving MHD is discussed in the Appendix \ref{A}.

To obtain the distribution of the non-thermal particles, we solve Parker's transport equation for non-thermal particles as follows,
\begin{equation}
\frac{\partial f}{\partial t}+(\mathbf{v}+\mathbf{v_d})\cdot\nabla f-\frac{1}{3}\nabla \cdot \mathbf{v} \frac{\partial f}{\partial \ln p_{\rm{NT}}} = \nabla\cdot(\boldsymbol{\kappa}\nabla f)  \label{distrib}
\end{equation}
where $f(x_{\rm{NT}},p_{\rm{NT}},t)$ is the distribution function, which is function of position, $x_{\rm{NT}}$, momentum, $p_{\rm{NT}}$, and time, $t$. $\mathbf{v_d}$, and $\boldsymbol{\kappa}$ are the particle drift velocity, and the spatial diffusion tensor, respectively. The Parker equation assumes a nearly isotropic momentum distribution. The spatial diffusion coefficient $\boldsymbol{\kappa}$ is determined by the properties of the turbulence, such as diffusion in MHD turbulence \citep{giacalone1999} or Bohm diffusion \citep{kang2007,hussein2014}.

Since we have assumed an nearly isotropic momentum distribution of $p_{\rm NT}$, the pressure of the non-thermal fluid is given by \citep{kudoh2016},
\begin{eqnarray}
    P_{\rm{NT}} = (\gamma_{\rm{NT}}-1)E_{\rm{NT}}, \label{pnt}
\end{eqnarray}
where $\gamma_{\rm{NT}}$ is the adiabatic index for the non-thermal fluid. $E_{\rm{NT}}$ is the internal energy density of the non-thermal fluid. 
The energy density of the non-thermal fluid at certain time and position is calculated by the collecting the second-order moment of the distribution function within the grid, 
\begin{eqnarray}
   E_{\rm{NT}} = \int E(p_{\rm NT}) f(p_{\rm NT}) dp_{\rm NT}^3,  \label{ent}
\end{eqnarray}
where $E(p_{\rm NT})$ is the energy of a non-thermal particle as a function of its momentum.

To solve Parker's transport equation, we solve its Fokker-Planck form expressed for $F=fp_{\rm{NT}}^2$ \citep{zhang1999,florinski2009,pei2010,kong2017,li2018},
\begin{equation}
\frac{\partial F}{\partial t} = -\nabla\cdot[(\nabla\cdot\boldsymbol{\kappa+\mathbf{v}})F]+\frac{\partial}{\partial p_{\rm{NT}}}\left[\frac{p_{\rm{NT}}}{3}\nabla\cdot\mathbf{v}F\right] + \nabla\cdot(\nabla\cdot(\boldsymbol{\kappa}F)). \label{sde}
\end{equation}
Here we only consider cases with $\mathbf{v_d} \ll \mathbf{u_{\rm{NT}}}$, which is a good approximation for solar flare reconnection (e.g., $\mathbf{v}_d \sim$ 0.01 km s$^{-1}$ for a 10 keV proton, current sheet thickness of 100 km, and a magnetic field of 100 G \citep{isenberg1979}).
The solution of this equation can be obtained by solving the corresponding stochastic differential equations (SDE) \citep{zhang1999,florinski2009}, the details of the method used to solve SDEs can be found in \citet{li2018}.

\subsection{Simulation Setup}
\subsubsection{Setup for Magnetic Reconnection Simulations}
\label{s.2.2.1}
For the initial condition, we adopt a thin, force-free current sheet within the simulation domain, which has dimensions of $L_x = 2$ and $L_y = 1$. The domain consists of a uniform Cartesian grid of $N_x\times N_y = 2048\times1024$ with $-1 < x <1$ and $-0.5 < y < 0.5$. The initial magnetic field 
\begin{equation}
\mathbf{B} = B_0 \tanh \left(\frac{y}{\lambda}\right)  \hat{x} + B_0 \cosh^{-1} \left(\frac{y}{\lambda}\right)  \hat{z}, \label{bini}
\end{equation}
where $B_0=1$ is the initial strength of the magnetic field, and $\lambda=0.005$ is half the thickness of the current sheet. The initial current sheet is resolved by about 10 cells. This simulation does not include any external guide field. The second term, similar to a guide field, is included to ensure force-free throughout the entire domain. The $x$ boundaries are periodic, and open boundaries are implemented for along the $y$ direction.

A magnetic perturbation is added to magnetic field for triggering magnetic reconnection, with the vector potential,
\begin{equation}
\phi_B(x,y) = \phi_0 B_0 \cos\left(\frac{2\pi y}{L_y}\right) \cos\left( \frac{2\pi x}{L_x} \right),
\end{equation}
where $\phi_0=10^{-4}$ is the amplitude of the magnetic perturbation. 
For solar flare environment \citep{li2018}, the normalized parameters are chosen for proton number density $n_{H}=1.2\times10^{10}\rm{cm^{-3}}$, length scale $L_0= 5000$ km, Alfvén velocity $v_A=v_0=$ 1000 km s$^{-1}$, and magnetic field $B_0=$ 50 G. The resistivity is chosen as $\eta=10^{-5}$, so the Lundquist number is given by $\mathcal{S}=L_x v_A/\eta = 10^{5}$. The plasma $\beta$ for the MHD fluid is $\beta=2P_g/B_0^2$ = 0.1. The timescale is $\tau_A=L_x/v_A$.

\subsubsection{Setup for SDE Simulations}
\label{s.2.2.2}

As shown in PIC simulations of magnetic reconnection \citep{Zhang2021,French2023,guo2024}, a fraction of the thermal particles are pumped into the high energy power-law tail via several injection processes. Since MHD-SDE does not include this injection for selecting non-thermal particles from the thermal pool, we initialize a population of nonthermal particles uniformly with the density a fraction of the plasma density.

PIC simulations have found that the injection fraction can be on the order of $10\%$ but depending on various plasma parameters \citep{Zhang2021,French2023}. We define the initial density ratio, which is the fraction of initial non-thermal particles density to the thermal plasma density, $R_i=\rho_{\rm{NT},i}/\rho_{0}$. For the three simulation cases, CASE1, CASE2, and CASE3, the values of $R_i$ are 0.01, 0.1, and 0.2, respectively.

In this study, we focus on proton acceleration in a solar flare environment. Our interest is in proton acceleration within the keV to MeV range, with the initial proton energy set to be 10 keV. In this range, protons are non-relativistic, so we use $\gamma_{\rm{NT}}=5/3$ for the non-thermal fluid. The typical gyroradius of a 10 keV proton is $\sim 100$ cm, which is much smaller than the length scale of the simulations.

We have tested how the numerical error depending on the number of pseudo-particles per cell, as the particles introduce statistical errors in the simulation (Appendix \ref{A1}). We find that in the reconnection simulation, the results are robust when we inject 100 pseudo-particles per cell or more. For all the magnetic reconnection simulations in this paper, we inject approximately $N_p\approx2\times10^8$ pseudo-particles, corresponding to $\sim 100$ pseudo-particles in each cell.

The spatial diffusion coefficient tensor can be expressed as,
\begin{equation}
\kappa_{ij} = \kappa_\perp \delta_{ij}-\frac{(\kappa_\perp-\kappa_\parallel)\mathbf{B}_i\mathbf{B}_j}{\mathbf{B}^2},
\end{equation}
where $\kappa_\perp$ and $\kappa_\parallel$ are the perpendicular and parallel diffusion coefficients.
We assume that isotropic Kolmogorov magnetic turbulence is well-developed and the gyroradii of the energetic protons are much smaller than the correlation length of the turbulence. Therefore, we can use the expression for the parallel spatial diffusion coefficient as follows \citep{giacalone1999},
\begin{equation}
\kappa_\parallel(v) = \frac{3v^3}{20L_c\Omega^2_0\sigma^2}\csc\left(\frac{3\pi}{5}\right)\left[1+\frac{72}{7}\left(\frac{\Omega_0L_c}{v}\right)^{5/3}\right],
\end{equation}
where $v$, $L_c$, $\Omega_0$, and $\sigma^2$ represent the particle speed, turbulence correlation length, particle gyrofrequency, and the normalized wave variance of turbulence, respectively. In these simulations, we adopt $L_c = L_0/3$ which is the largest
eddy size in a reconnection-driven turbulence \citep{huang2016}, and $\sigma^2=\left<\delta B^2\right>/B_0^2=1$. The normalization of the spatial diffusion coefficient is given by $\kappa_0=L_0v_A=5\times10^{16}\rm{cm}^2\rm{s}^{-1}$. Hence, $\kappa_\parallel = 0.003\kappa_0$ for 10keV protons and scale with particle momentum $p^{4/3}$. We adopt the perpendicular spatial coefficient as $\kappa_\perp = 0.01\kappa_\parallel$.

\section{Simulation Results}
\label{s3}

\begin{figure}[ht!]
\includegraphics[width=\textwidth]{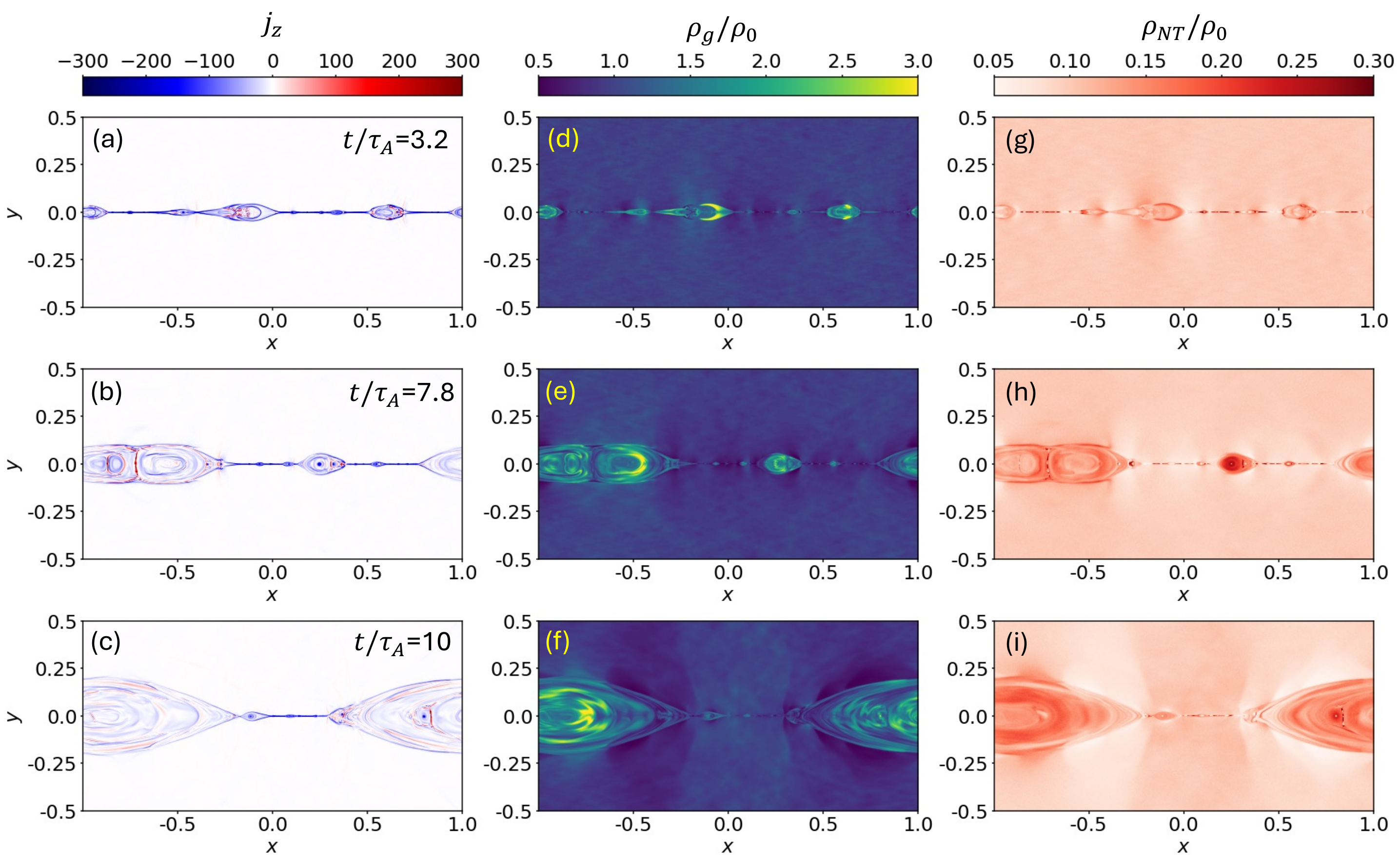}
\caption{Left panels: The distribution of out-of-plane current density, $j_z$, at (a) $t/\tau_A=3.2$, (b) $7.8$, and (c) $10$ for the CASE2 ($R_i = \rho_{\rm{NT},i}/\rho_{0} = 0.1)$, respectively. Middle panels: The distribution of plasma density, $\rho_g$, at (d) $t/\tau_A=3.2$, (e) $7.8$, and (f) $10$ for the same case. Right panels: The distribution of non-thermal particle density, $\rho_{\rm{NT}}$, at (g) $t/\tau_A=3.2$, (h) $7.8$, and (i) $10$ for the same case. \label{f1}}
\end{figure}

\begin{figure}[ht!]
\includegraphics[width=\textwidth]{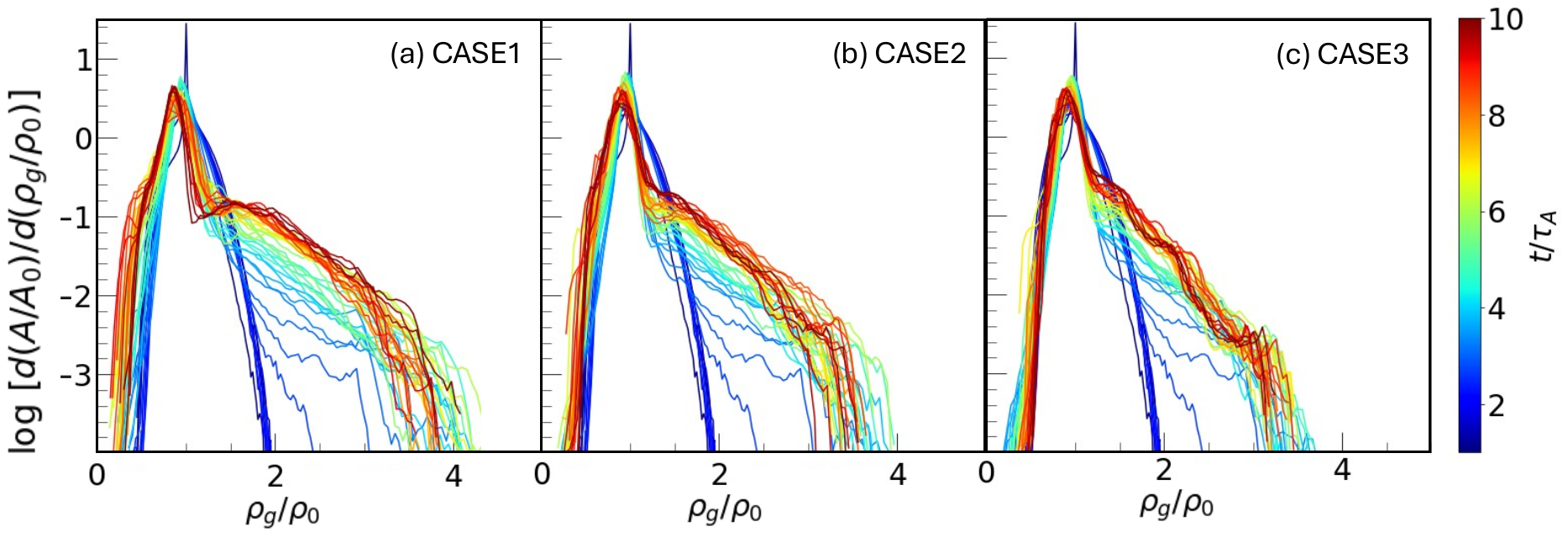}
\caption{Time evolution of the plasma density probability distributions function for the (a) CASE1, (b) CASE2, and (c) CASE3. The plasma density is normalized by the initial values in each simulation. \label{f2}}
\end{figure}

The initial current sheet keeps thinning under the influence of the perturbation, leading to magnetic reconnection and the generation of magnetic islands \citep{loureiro2007, bhattacharjee2009, comisso2016, li2018}. In Figure \ref{f1}, we present the time evolution of the out-of-plane current density $j_z$, plasma gas density $\rho_g$, and non-thermal particle density $\rho_{\rm{NT}}$ for $t/\tau_A=3.2, 7.8$, and 10, respectively for the CASE2 ($R_i=0.1$). Magnetic reconnection generates outflows with speeds on the order of the upstream Alfvén speed. As shown in the middle panels of Figure \ref{f1}, small magnetic islands continuously form and merge along the unstable current sheet, leading to the continuous growth of the largest magnetic island size. These magnetic islands contract due to magnetic tension forces, resulting in the generation of high-density regions inside them. The reconnection layer is highly compressible.

\begin{figure}
\includegraphics[width=\textwidth]{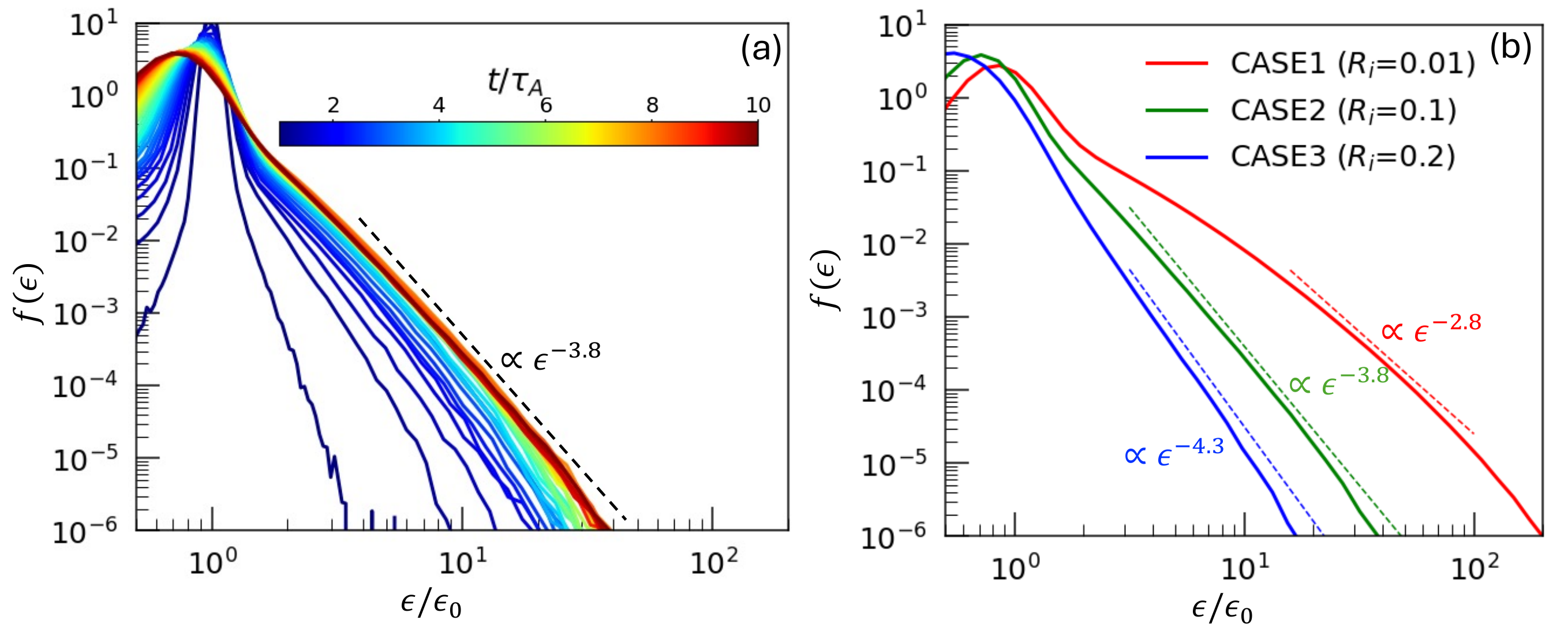}
\centering
\caption{(a) The time evolution of the energy spectrum of non-thermal protons across the entire domain for the CASE2. Colors indicate different time steps. Dashed lines indicate the slope of the saturated spectrum. (b) The energy spectra of three cases at $t/\tau_A=10$ for CASE1 (red), CASE2 (green), and CASE3 (blue). Dashed lines indicate the slope of the power-law. \label{f3}}
\end{figure}

The density of the thermal plasma changes adiabatically, while the nonthermal component undergoes additional spatial diffusion. As a result, as shown in the middle and right panels of Figure \ref{f1}, the high-density regions of non-thermal particles do not exactly match the plasma density distribution.
While we do not present it here, the distribution of the non-thermal particles strongly depends on the spatial diffusion model. When we adopt isotropic spatial diffusion ($\kappa_\perp=\kappa_\parallel$), the structures in the non-thermal particle density are smeared out.

\begin{figure}[ht!]
\includegraphics[width=\textwidth]{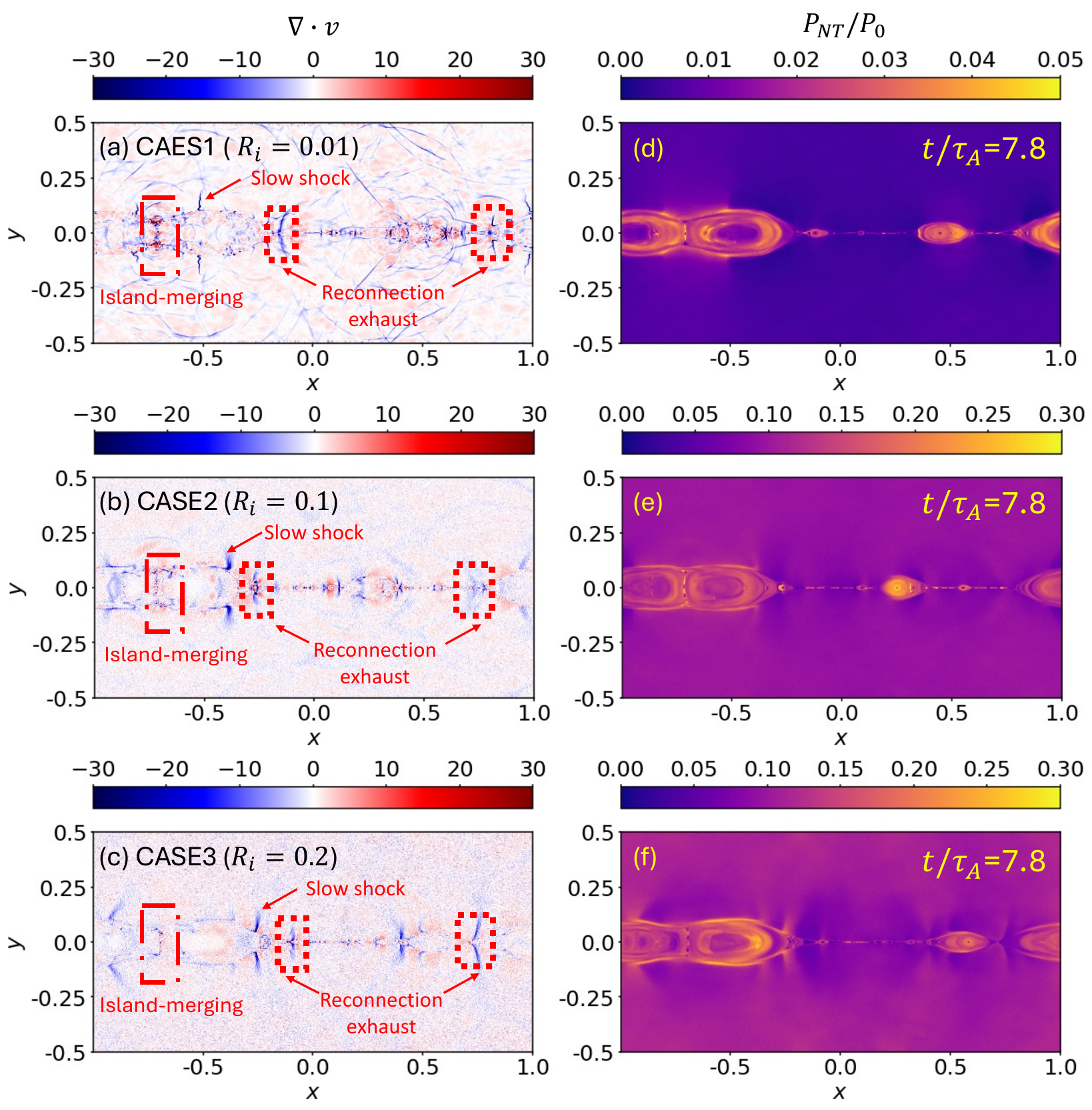}
\caption{Left panels: The distribution of non-thermal fluid $\nabla\cdot\mathbf{v}$ for the (a) CASE1, (b) CASE2, and (c) CASE3 at $t/\tau_A=7.8$. The regions indicating reconnection exhaust and island merging are denoted by dotted, and dot-dashed red boxes. The merging of the magnetic islands induces pressure disturbances in the upstream region, which eventually evolve into slow shocks. (Right panels) The distribution of pressure for the (d) CASE1, (e) CASE2, and (f) CASE3 at $t/\tau_A=7.8$. Note that the color bar range is adjusted for (d) due to the dynamic range of $P_{\rm{NT}}$.
\label{f4}}
\end{figure}

We present the probability distribution function (PDF) of plasma density in Figure \ref{f2} for all three cases. We find that highly compressed regions, $\rho_g/\rho_0>1.5$, are generated during magnetic reconnection. As the fraction of energetic particles increases, the area of highly compressed region is reduced. While the compression region leads to particle acceleration, it is clear that pressure from energetic particles mediates the compressibility of the reconnection layer.

\begin{figure}[ht!]
\centering
\includegraphics[width=0.5\textwidth]{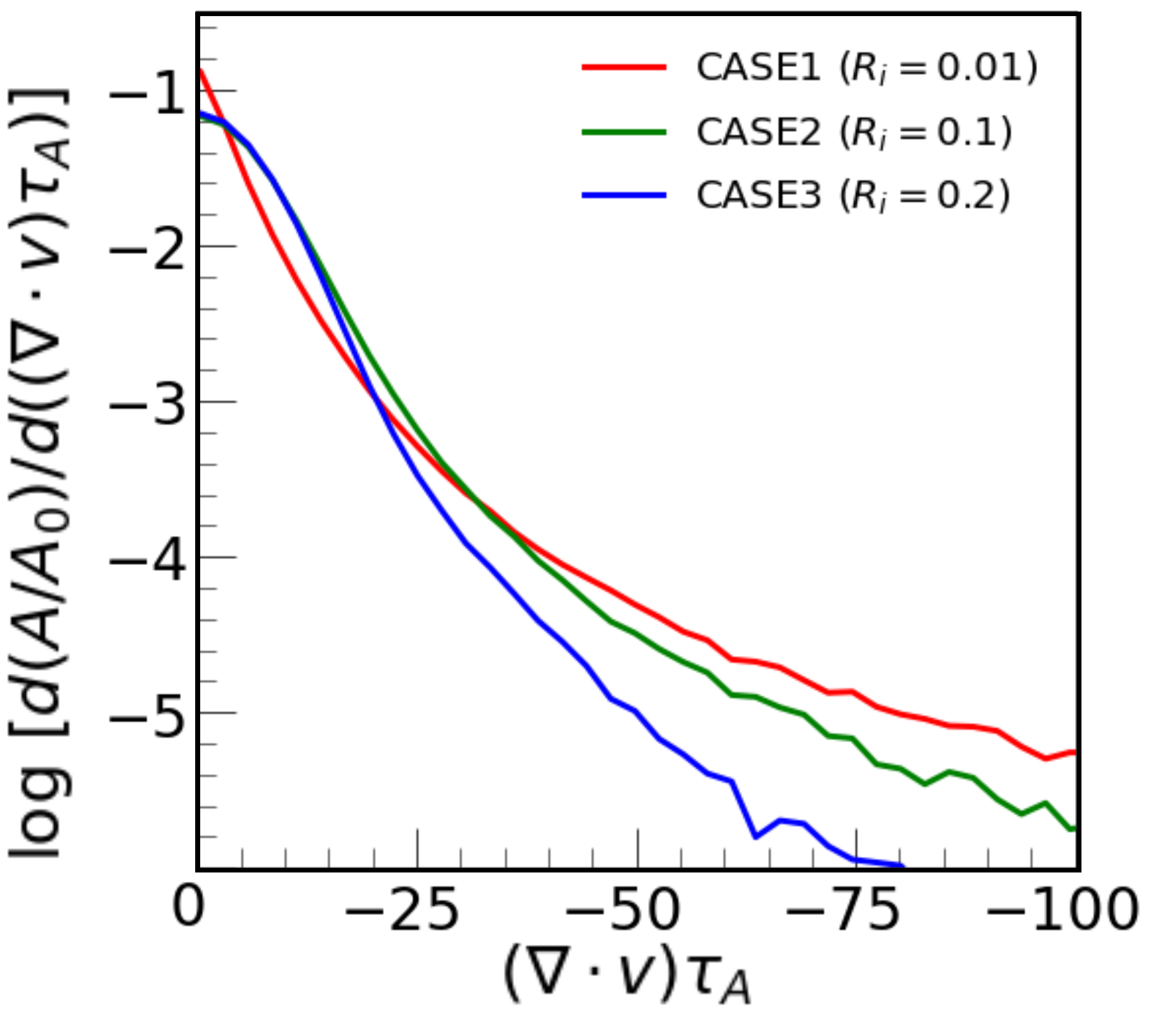}
\caption{The probability distribution function of compression, $\nabla\cdot\mathbf{v}$, is averaged within $5\le t/\tau_A \le10$ across the entire domain. The results are shown for the CASE1 (red), CASE2 (green), and CASE3 (blue). $A$ is the area corresponding to each compression bin, and $A_0$ is the total domain size.  \label{f5}}
\end{figure}

Figure \ref{f3} (a) shows the time evolution of the non-thermal proton energy spectrum for CASE2. A clear power-law spectrum quickly emerges and extends to high energy. In the late stage, the spectral slope converges to roughly ``-4'' when reconnection becomes saturated. We have not included the spectral evolution for the other cases, but they have the same trend. Figure \ref{f3} (b) presents the energy spectra for cases with different fraction of non-thermal particles. All cases show sustainable particle acceleration and a clear power-law spectrum. We find that larger fraction of non-thermal particles results in a steeper energy spectrum. For the CASE1, 2, and 3, their energy spectra follow a power-law with $f(\epsilon)\propto \epsilon^{-2.8}$, $\epsilon^{-3.8}$, and $\epsilon^{-4.3}$, respectively. The slope in the smallest fraction of non-thermal particle case (CASE1) is slightly steeper than in the case without feedback (not shown). This highlights that feedback is crucial for determining the slope of the energy spectrum.
While we have not included here, the magnitude of the diffusion coefficient also controls the efficiency of particle acceleration and spectral index, similar to \citet{li2018}.

According to the Parker equation, particles are accelerated by $\nabla\cdot\mathbf{v}$, which allows for better tracking of the acceleration compared to the density distribution. Figure \ref{f4} shows divergence of the plasma flow $\nabla\cdot\mathbf{v}$ and the non-thermal pressure $P_{\rm{NT}}$ for all cases at $t/\tau_A =7.8$. \citet{li2018} suggested that Fermi first-order acceleration occurs in regions with strong compression, such as the reconnection exhaust region and magnetic island merging region. These regions are indicated in Figure \ref{f4} (a), (b), and (c). In addition, slow shocks are generated by pressure disturbances due to the merging of magnetic islands \citep{zenitani2010,arnold2022}. In CASE1, clear compression regions are observed. These regions are considered as important regions for particle acceleration \citep{li2018}. $P_{\rm{NT}}$ increases with an increasing fraction of non-thermal particles, as shown in the right panels of Figure \ref{f4}. Correspondingly, the compressed regions, including the magnetic island merging and reconnection exhaust regions, become less compressible, seen in the left panels of Figure \ref{f4}. This can be more clearly illustrated in Figure \ref{f5}. 
As shown in the right panels of Figure \ref{f4}, $P_{\rm{NT}}$ is generally enhanced by density enhancement due to converging flows and particle acceleration due to compression. Active particle acceleration occurs along the current sheet, island merging region, reconnection exhaust, and slow shock region, so $P_{\rm{NT}}$ is enhanced in these regions. 
When the high $P_{\rm{NT}}$ is generated, a force acts from the high-pressure region to the low-pressure region (see Equations (\ref{meq})). Thus, the gradient of $P_{\rm{NT}}$ leads to feedback in momentum and energy; in other words, feedback from non-thermal particles mediates the compressibility of the reconnection layer.

Quantitatively, there are also differences in the distribution of velocity compression $\nabla\cdot\mathbf{v}$ for different fraction of energetic particles. Figure \ref{f5} shows the averaged PDF for $\nabla\cdot\mathbf{v}$, which measures how compressible the reconnection layer is. Since reconnection exhaust and island merging do not occur continuously, we averaged the PDFs of 20 snapshots within $5\le t/\tau_A \le 10$. Figure \ref{f5} shows less compression for the higher $R_i$ cases at the high compression region where $(\nabla\cdot\mathbf{v})\tau_A\lesssim -30$. Therefore, it is clear that feedback from non-thermal particles mediates pressure, leads to less compression, and consequently generates a steeper energy spectrum.

\section{Discussion and Summary}
To investigate astrophysical and space plasma processes that involve a large fraction of energetic particles, we have developed a new method for including feedback from non-thermal particles called MHD-SDE method. To obtain the distribution of non-thermal component during magnetic reconnection, we solve Parker's transport equation  via its corresponding stochastic differential equations. While similar particle acceleration simulations have been conducted \citep{li2018}, here  the pressure of non-thermal particles is incorporated into the MHD equations to include its feedback.

The extended MHD equations are solved using the HOW-MHD code \citep{seo2023}, providing high-order accuracy. 
As a first application, we use this method to study particle acceleration during low $\beta$ magnetic reconnection. To demonstrate the role of feedback from non-thermal particles, we simulate three cases with non-thermal particles at 1\%, 10\%, and 20\% of the total plasma density, respectively. We find that magnetic reconnection accelerates a large fraction of non-thermal particles to high energy. The feedback from non-thermal fluid mediates the compressibility of the reconnection layer. The pressure of non-thermal fluid suppresses the generation of highly compressed structures, leading to reduced energy gain in particle acceleration process. Consequently, the energy spectrum of non-thermal protons exhibits a steeper power law when the non-thermal fraction is larger. Therefore, it is essential to consider the feedback from non-thermal particles, as there is a considerable difference in the energy spectrum slope with or without feedback \citep{li2018}.

The feedback effect is different from the findings by \citet{drake2019,arnold2021,Yin2024}, who suggest the development of pressure anisotropy in the energetic particles leads to a reduction in magnetic tension that drives energy release. However, we observed a strong feedback effect even when the momentum distribution is assumed to be nearly isotropic. The feedback mechanism in our reconnection simulation is due to the interplay between the energetic particle pressure and compression structures.
In further studies, we plan to solve the focused transport equation \citep{zank2014,leRoux2015,zhang2017,kong2022} to include the anisotropic momentum distribution of energetic particles.

In the following study, we will investigate the effect of the feedback of non-thermal particles in the magnetic reconnection region in realistic solar flare magnetic geometry. Our goal is to apply this study to understand the acceleration and distribution of energetic particles in shocks, turbulence, and magnetic reconnection occurring within the various space and astrophysical environments. 
\label{s4}

\section*{Acknowledgements}
We acknowledge the support from Los Alamos National Laboratory through the LDRD program, DOE OFES, and NASA programs through grant 80HQTR21T0005, 80HQTR21T0087, 80HQTR21T0117, NNH230B17A, NNH240B72A, NNH24OB107, and the ATP program. F. G. acknowledges NSF Award No. 2308091. X.L. acknowledges the support from NASA through Grant 80NSSC21K1313, National Science Foundation Grant No. AST-2107745, Smithsonian Astrophysical Observatory through subcontract No. SV1-21012, and Los Alamos National Laboratory through subcontract No. 622828.  The simulations used resources provided by the Los Alamos National Laboratory Institutional Computing
Program, the National Energy Research Scientific Computing Center (NERSC) and the Texas Advanced Computing
Center (TACC). 

\appendix
\section{Schemes for simulations}
\label{A}
To solve the MHD equations, we utilize the HOW-MHD code \citep{seo2023}, which employs finite difference fifth-order Weighted Essentially Non-Oscillatory (WENO) reconstruction and a fourth-order Strong-Stability-Preserving Runge-Kutta (SSPRK) method with a high-order constrained flux algorithm. This code has a high effective resolution, allowing it to provide detailed non-linear structures such as shocks and turbulence, which are important for tracking the acceleration process at sharp MHD structures. To achieve a high-order accuracy in this code, we adopt a high-order difference method to obtain divergence and non-thermal fluid pressure \citep{romao2012},

\begin{equation}
\frac{\partial Q}{\partial x} = \frac{aQ_{i+3} - bQ_{i+2} + cQ_{i+1} - cQ_{i-1} + bQ_{i-2} - aQ_{i-3}}{d\Delta x}, \label{hd}
\end{equation}

\noindent where $Q$ is an arbitrary function and $i$ is the number of the grid along the $x$ direction. In Equation (\ref{hd}), we omitted the $y$ and $z$ directions for simplification. For a fourth-order difference, $a=0$, $b=1$, $c=8$, and $d=12$, and for a sixth-order difference, $a=1$, $b=9$, $c=45$, and $d=60$. As shown in the pressure balance test, this method suppresses spurious oscillations (see section \ref{A4}).

\section{Test problems}
\label{B}
\begin{figure}[ht!]
\plotone{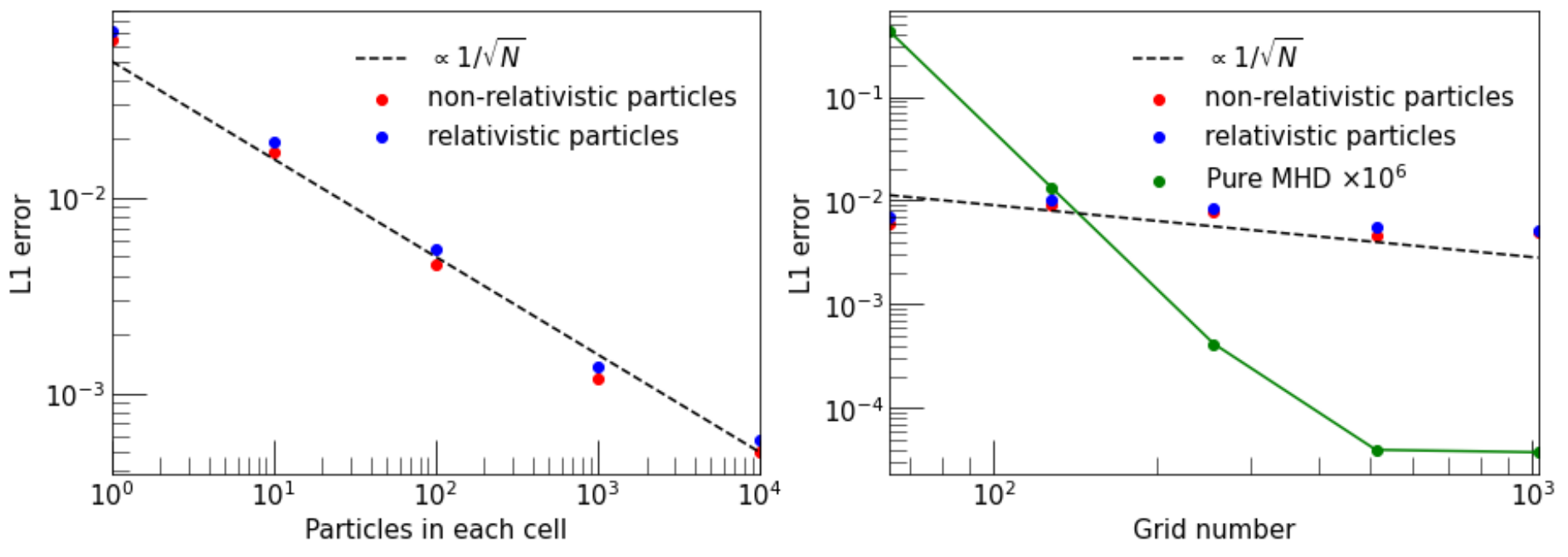}
\caption{(Left) L1 error dependency on the number of particles per cell. The red and blue dots indicate particles in the non-relativistic and relativistic regimes, respectively. The black dashed line indicates the inverse proportionality to the square root of the total number of particles. (Right) L1 error dependency on grid resolution. The color code is the same as in the left panel. The green solid line indicates the L1 error of pure MHD simulation.  \label{Af1}}
\end{figure}

\begin{figure}[ht!]
\plotone{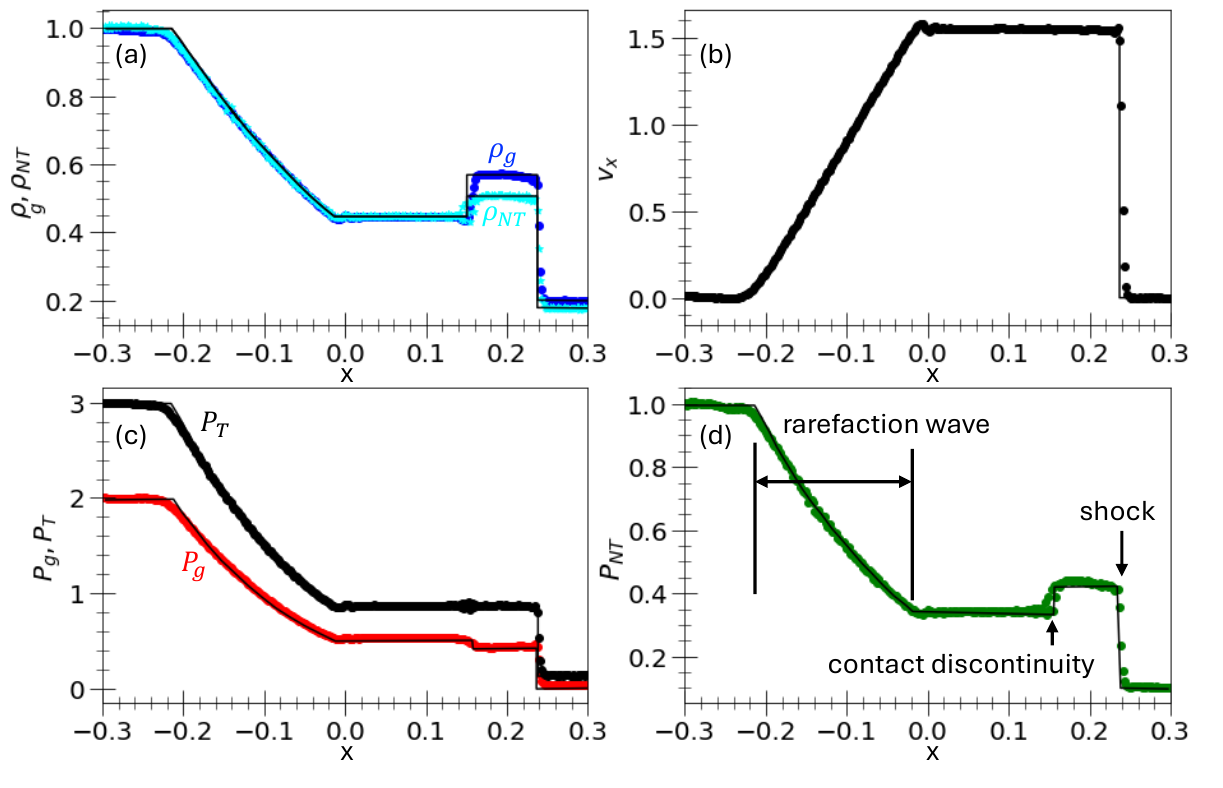}
\caption{2D hydrodynamic shock tube with relativistic non-thermal fluid simulation results for (a) density of the plasma (blue dot) and non-thermal fluid (cyan dot), (b) $v_x$, (c) thermal plasma pressure (red dot) and total pressure (black dot), and (d) non-thermal fluid pressure at $t$=0.1. Solid black lines indicate the analytic Riemann solution. \label{Af2}}
\end{figure}

To ensure proper demonstration of the non-thermal fluid pressure obtained from the Parker equation and its adequate feedback in the hydrodynamic regime, we benchmark seveal test problem including the shock tube test using non-thermal fluid pressures as suggested in \citet{kudoh2016}. We denote primitive variables as $\mathbf{u}=(\rho_g,v_x,P_g,P_{\rm{NT}})$ for the hydrodynamic shock tube test in \ref{A2} and the pressure balance mode in \ref{A4}, and $\mathbf{u}=(\rho_g,v_x,v_y,v_z,B_x,B_y,B_z,P_g,P_{\rm{NT}})$ for the magnetohydrodynamic shock tube test in section \ref{A3} and reflection shock test in section \ref{A5}. All tests adopt a Courant-Friedrichs-Lewy number of 1.5, leveraging the advantages of SSPRK \citep{seo2023}. The MHD time step, $\Delta t$, is determined by the maximum eigenvalues within the domain and grid resolution \citep[see][for details]{seo2023}.

\subsection{Advection of entropy wave}
\label{A1}
Solving SDE involves using the Lagrangian method, which generates random errors. To estimate the effect of this error, we conduct an advection of entropy wave test. This test is performed in 1D, with density perturbation given by,
\begin{equation}
\rho_g = \rho_0 + \epsilon_\rho \sin \left(\frac{2\pi}{L}\right),
\end{equation}
where $L=1$ is the domain size, $\rho_0=1$, and $\epsilon_\rho=0.1$. The other parameters are set as follows: $P_g=1$, $P_{\rm{NT}}=1$, and $v_x = 10$, with no magnetic field and no spatial diffusion, $\kappa_\perp=\kappa_\parallel=0$. The L1 error is determined as,
\begin{equation}
\text{L1 error}= \sqrt{\sum_{s=1}^5\left(\sum_i\frac{|\mathbf{u}_i^s(t)-\mathbf{u}_i^s(0)|}{N}\right)},
\end{equation}
where $N$ is the number of grid points, $i$ is the cell position, and $s$ is the component of the primitive variables. For the test on the dependency of the number of particles per cell, we use 512 grid points, and for the test on the dependency of the grid number, we put 100 particles in each cell. As shown in the left panel of Figure \ref{Af1}, the random error decreases with the square root of the number of particles per cell, $\sqrt{N_p}$, as expected. The L1 error for 1000 particles per cell is approximately $1\%$ of the density perturbation. In this limit, the random error has a marginal effect on the wave. Hence, while a low number of particles can distort the hydrodynamic structures, a sufficiently large number of particles can minimize this effect. The L1 error of pure MHD is much less than the error from SDE (as shown by the solid green line in the right panel of Figure \ref{Af1}; this error saturates in the double-precision limit). Increasing the grid number increases the total number of particles, so the L1 error also decreases with $\sqrt{N_p}$. This pattern indicates that the random error when using the Lagrangian method to solve SDE is much larger than those caused by numerical dissipation.

\subsection{Hydrodynamic shock}
\label{A2}
The first test is a simple hydrodynamic shock tube, initially containing high-density and high-pressure fluid on one side, and low-density and low-pressure fluid on the other. The initial conditions are specified as follows: for the left side, $\mathbf{u}_L=(1.0,0.0,2.0,1.0)$, and for the right side, $\mathbf{u}_R=(0.2,0.0,0.02,P_{\rm{NT,2}})$. Where $P_{\rm{NT,2}}=0.1$ for relativistic non-thermal fluid and $P_{\rm{NT,2}}=0.06$ for non-relativistic non-thermal fluid.
The initial density of the non-thermal fluid is determined by the adiabatic condition, $\rho_{\rm{NT}} = \rho_{\rm{NT},0} P_{\rm{NT}}^{1/\gamma_{\rm{NT}}}$, where $\rho_{\rm{NT},0}=1$ for this simulation. In this test, we adopt relativistic non-thermal fluid, with the pressure given by Equation (\ref{ent}). The domain is $[-0.5, 0.5]\times[-0.25, 0.25]$ with a grid resolution of $512\times256$. We inject 1,000 pseudo-particles into each cell to solve the SDE. The test results are presented in Figure \ref{Af2} at $t$=0.1. In this figure, we averaged the primitive variables along the $y$-axis to reduce the random noise. In this test, spatial diffusion is not included to allow for direct comparison with the shock tube test of \citet{kudoh2016}, but we have adopted spatial diffusion in the MHD shock tube (section \ref{A3}) and reflecting shock (section \ref{A5}).

As shown in Figure \ref{Af2}, an expansion wave ($x$=-0.23 $\sim$ -0.02), a contact discontinuity ($x$=0.14), and a shock ($x$=0.24) are generated. Thanks to the high-order WENO Reconstruction in the HOW-MHD code, the contact discontinuity and shock are resolved within approximately 3-4 cells. The non-thermal fluid pressure, $P_{\rm{NT}}$, is obtained by solving Equation (\ref{sde}), as illustrated by the green dot in the bottom-right plot of Figure \ref{Af2}. The solution is well aligned with the analytic Riemann solution provided in \citet{kudoh2016}. This result demonstrates that solving the SDE can provide proper non-thermal fluid pressure and adequate feedback in the hydrodynamic regime. At the contact discontinuity, small spurious oscillations are generated, resulting in approximately a 2\% error in total pressure. This error is primarily due to the numerical dissipation of the non-thermal fluid (see section \ref{A4}).

\begin{figure}[ht!]
\plotone{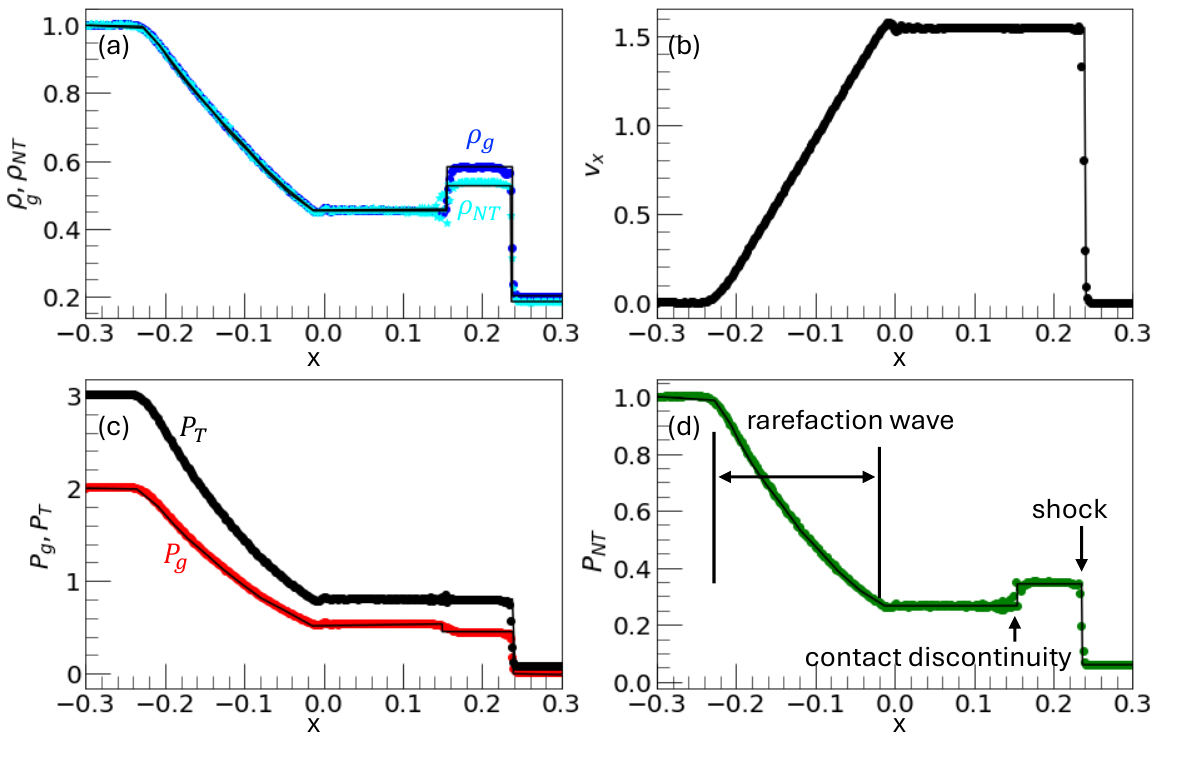}
\caption{The hydrodynamic shock tube with non-relativistic non-thermal fluid simulation results for (a) density of the plasma (blue dot) and non-thermal fluid (cyan dot), (b) $v_x$, (c) thermal plasma pressure (red dot) and total pressure (black dot), and (d) non-thermal fluid pressure at $t$=0.1. Solid black lines indicate the analytic Riemann solution. \label{Af3}}
\end{figure}

We performed an additional test for the non-relativistic non-thermal fluid. The initial conditions are the same as in the first simulation, but the initial density of the non-thermal fluid is determined by the adiabatic condition with $\gamma_{\rm{NT}}=5/3$. Consequently, the initial density of the non-thermal fluid on the right side is larger than in the previous test. The test results are presented in Figure \ref{Af3}.

For the non-relativistic non-thermal fluid, the reduction in non-thermal fluid energy ($E_{\rm{NT}}=P_{\rm{NT}}/(\gamma_{\rm{NT}}-1)$) results in a reduced compression at the shock. This leads to a decrease in the jump of the plasma density, pressure, and velocity of the downstream of the shock (see $x$=0.2-0.3 in Figures \ref{Af2} and \ref{Af3}). However, since the non-thermal fluid pressure is given as $P_{\rm{NT}}=\rho_{\rm{NT}}^{\gamma_{\rm{NT}}}$, the non-thermal fluid pressure in the shock downstream region is larger than in the relativistic case, even though the compression is reduced. This test demonstrates that our method also works well within the non-relativistic non-thermal fluid limit.

\subsection{Magnetohydrodynamic shock}
\label{A3}

\begin{figure}
\plotone{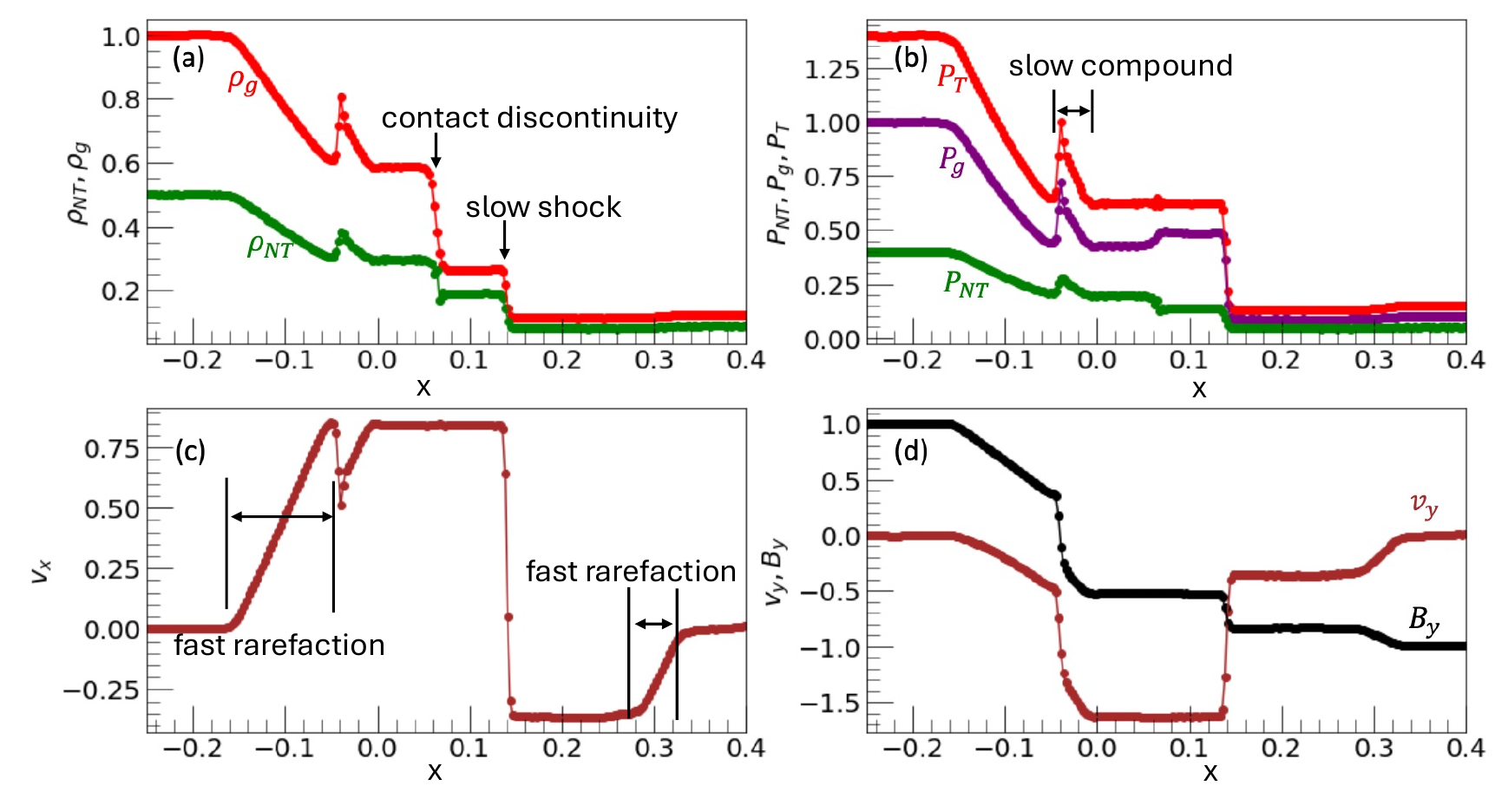}
\caption{The magentohydrodynamic shock tube with relativistic non-thermal fluid simulation results for (a) density of the plasma (red dot) and non-thermal fluid (green dot), (b) total pressure (red dot), thermal plasma pressure (purple dot), and non-thermal fluid pressure (green dot), (c) $v_x$, and (d) non-thermal fluid pressure at $t$=0.08. \label{Af4}}
\end{figure}

To use this method in MHD simulations, we conduct the Brio-Wu shock tube test \citep{brio1988} with relativistic non-thermal fluid, following the approach outlined in \citet{kudoh2016}. In all MHD test simulation, $\eta=0$. The initial conditions for the left side are given as $\mathbf{u}_L=(1.0,0.0,0.0,0.0,1.0,1.0,0.0,1.0,0.4)$, and for the right side, $\mathbf{u}_R=(0.125,0.0,0.0,0.0,1.0,-1.0,0.0,0.1,0.04)$. The simulation end time is $t=0.08$. The domain size and the number of injection pseudo-particles are the same as in the hydrodynamic shock tube test. Figure \ref{Af4} shows that this test generates fast rarefaction, slow compound, contact discontinuity, slow shock, and fast rarefaction, from left to right. For the first approach, we do not adopt spatial diffusion. As shown in Figure \ref{Af4}, our result agrees well with the result of \citet{kudoh2016}, for instance, the position and amplitude of the MHD structures (see Figure 11 in \citet{kudoh2016}). Especially, our code captures the contact discontinuity within 3-4 cells.

\begin{figure}[ht!]
\plotone{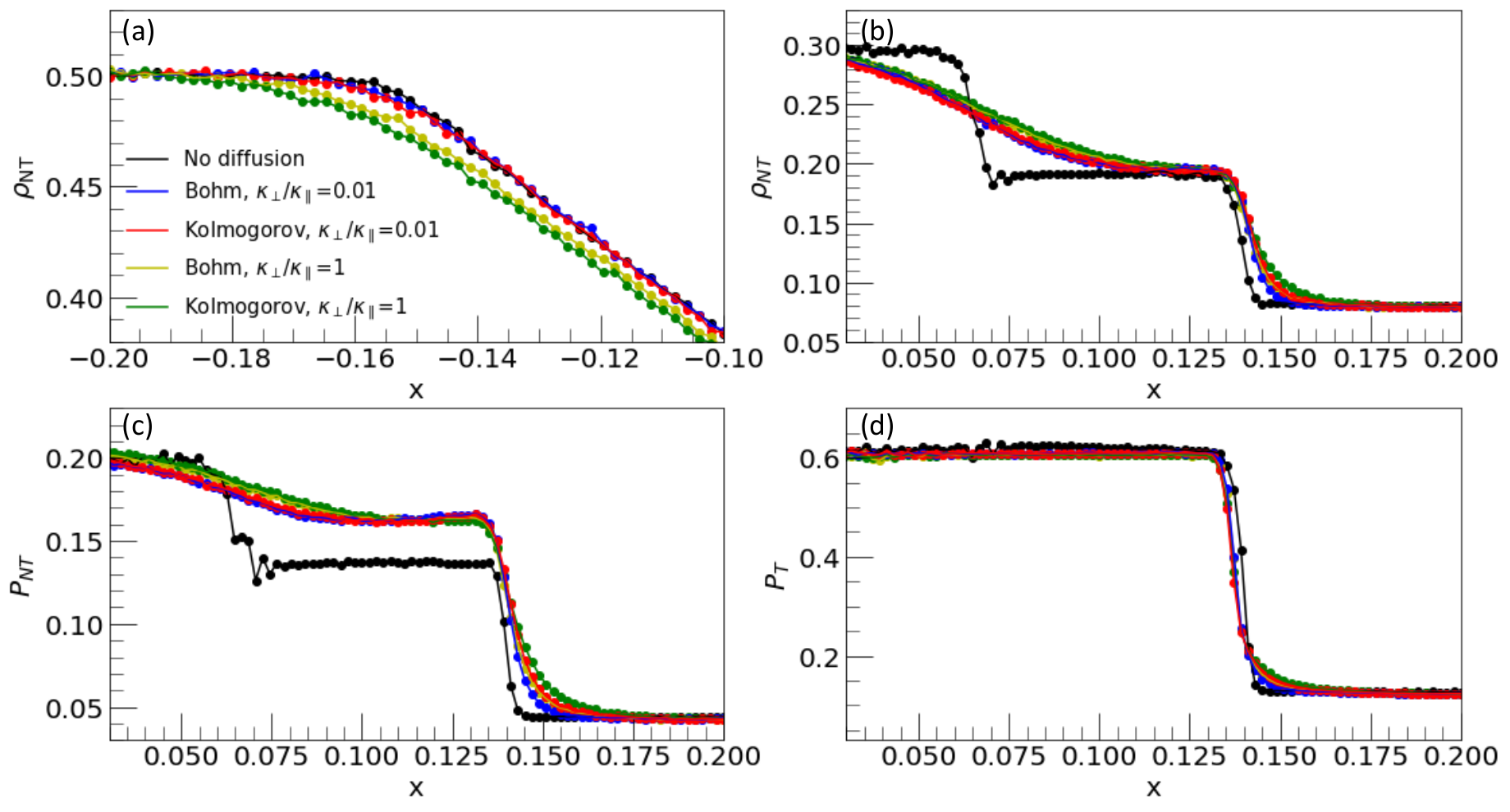}
\caption{The magnetohydrodynamic shock tube with relativistic non-thermal fluid simulation yields results at $t=0.08$ for (a) non-thermal fluid density in the rarefaction range, (b) non-thermal fluid density in the contact discontinuity and slow shock, (c) non-thermal fluid pressure in the same range, and (d) total pressure in the same range. The color code depends on the $\kappa$ model and spatial diffusion coefficient as follows: no spatial diffusion (black), Bohm diffusion with $\kappa_\perp/\kappa_\parallel=0.01$ (blue), diffusion in MHD turbulence with $\kappa_\perp/\kappa_\parallel=0.01$ (red), Bohm diffusion with $\kappa_\perp/\kappa_\parallel=1$ (yellow), and diffusion in MHD turbulence with $\kappa_\perp/\kappa_\parallel=1$ (green).} \label{Af5}
\end{figure}

For the next test, we incorporate spatial diffusion using two $\kappa$ models. The first model is the Bohm-like diffusion model, $\kappa_{\parallel} = \kappa_0(p_{\rm{NT}}/p_0)$ \citep{kang2007}, and the second model is diffusion in MHD turbulence model, $\kappa_{\parallel} = \kappa_0 (p_{\rm{NT}}/p_0)^{4/3}$ \citep{li2018}, where $p_0$ is the initial momentum of the pseudo-particle. Additionally, we examine the $\kappa_{\perp}/\kappa_{\parallel}$ dependence. In this simulation, we set $\kappa_0=0.003$. We present the diffusion dependence test in Figure \ref{Af5}, with a particular focus on the right fast rarefaction (panel (a) in Figure \ref{Af5}), contact discontinuity, and slow shock (panels (b)-(d) in Figure \ref{Af5}).

As expected from the model, the spatial diffusion coefficient is larger for the MHD turbulence model at the same momentum. Consequently, more diffusion of the structures is observed in the MHD turbulence model for rarefaction and discontinuity than in the Bohm diffusion model, but the difference is marginal. With non-zero $B_y$ (see black line in Figure \ref{Af4} (d)), a lower $\kappa_{\perp}/\kappa_{\parallel}$ leads to a smaller $\kappa_{xx}$ and, consequently, less spatial diffusion when $\kappa_{\perp}/\kappa_{\parallel}=0.01$ than when it is 1. A $\kappa_{\perp}/\kappa_{\parallel}$ value of 1 indicates significant diffusion of non-thermal fluid perpendicular to the magnetic field direction, leading to increased diffusion in both structures. Compared to the rarefaction, spatial diffusion of non-thermal fluid does not smooth the shock as much because of its large compression.

A contact discontinuity is noticeable around $x=0.07$ in panels (b)-(d) of Figure \ref{Af5}. As mentioned earlier, without spatial diffusion, spurious oscillations occur around the contact discontinuity. However, when including spatial diffusion, these oscillations disappear, and sharp boundaries become continuous structures. Due to the diffusion of these particles, non-thermal fluid density and pressure are slightly increased. However, total pressure is preserved, as shown in panel (d) of Figure \ref{Af5}. Since the shock Mach number is determined by the total pressure jump, spatial diffusion does not affect the strength of the shock.

In summary, spatial diffusion of the non-thermal fluid can smooth out MHD structures, such as contact discontinuities or rarefactions, but it only has a marginal effect that cannot significantly distort the MHD structure.

\begin{figure}
\plotone{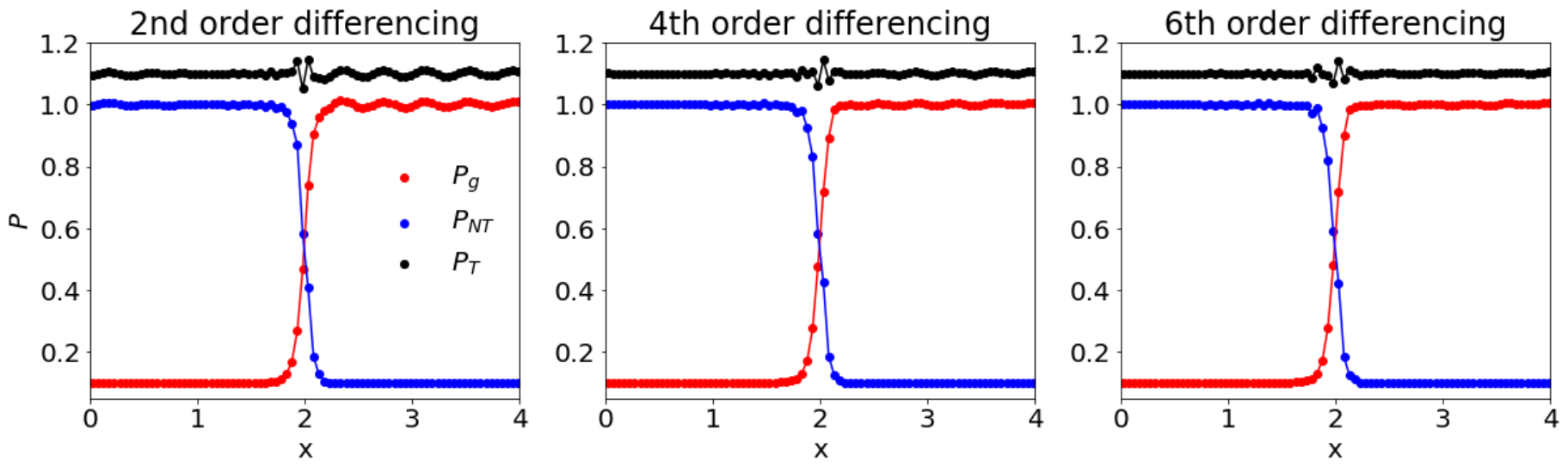}
\caption{The Advection of the Pressure Balance Mode test at $t=2$ with (left) 2nd, (middle) 4th, and (right) 6th order differencing method (Equation (\ref{hd})). Red dot indicates the plasma pressure profile, blue dot indicates the non-thermal fluid pressure, and black dot indicates the total pressure.\label{Af6}}
\end{figure}

\begin{figure}
\plotone{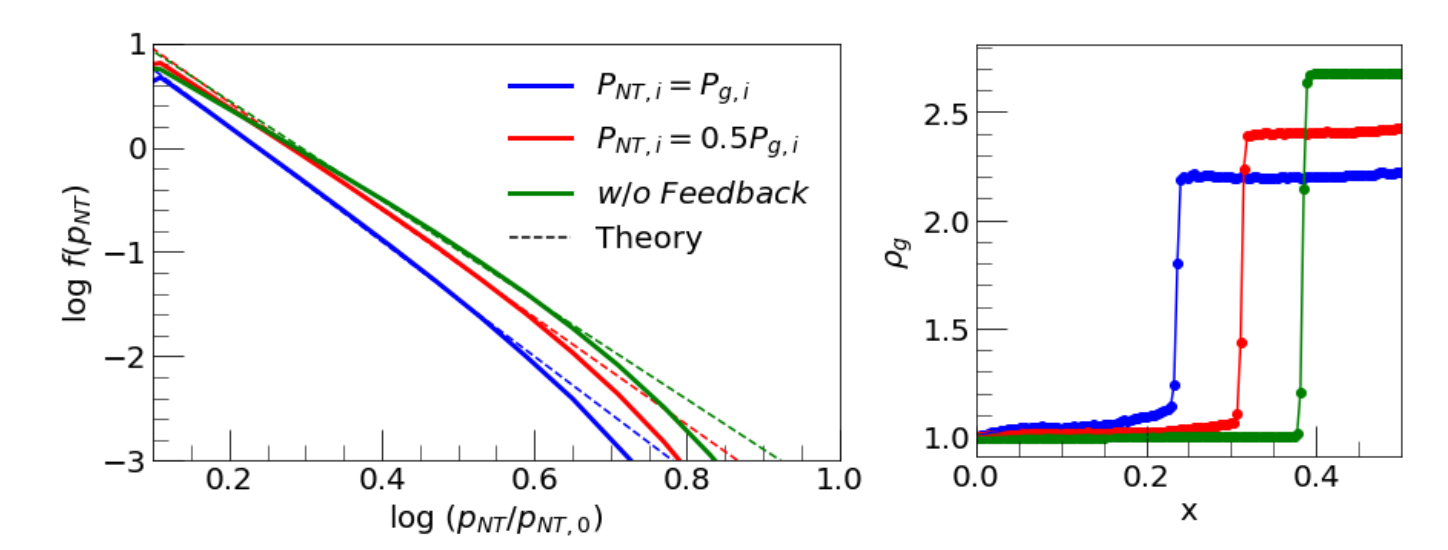}
\caption{(Left) The energy spectrum of non-thermal fluid at $t=0.5$ with $P_{\rm{NT},i}=1P_{g,i}$ (blue solid), $P_{\rm{NT},i}=0.5P_{g,i}$ (red solid), and $P_{\rm{NT},i}=0P_{g,i}$ (green solid). The dashed lines represent the analytic solutions for the respective shock compression ratios. (Right) The density profile around the shock region at $t=0.5$. The color code used is the same as in the left panel.  \label{Af7}}
\end{figure}

\begin{figure}[ht!]
\plotone{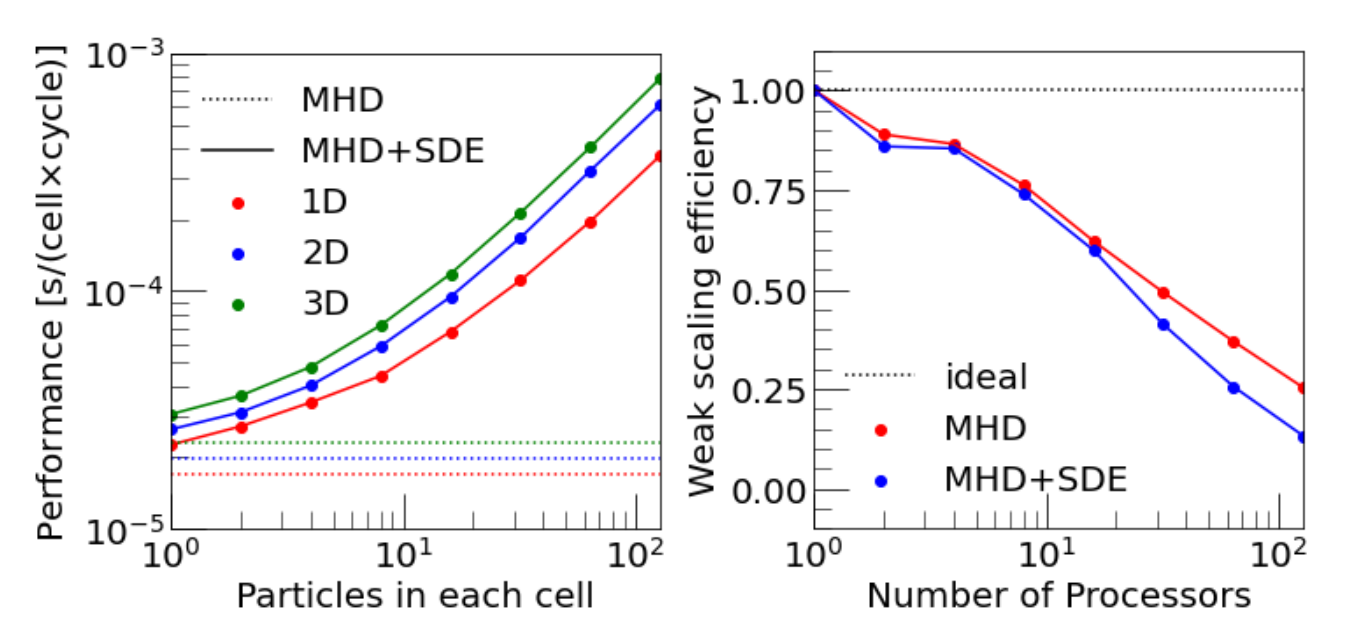}
\caption{(Left) Computational performance versus the number of particles per cell, which is defined as the computational time divided by the total number of cells and the number of MHD cycles. Red, blue, and green colors indicate tests performed in 1D, 2D, and 3D, respectively. The dashed lines represent the performance with only solving MHD, with the dimension indicated by the respective color code. (Right) Weak scaling comparison between only MHD (red) and MHD with the SDE method (blue).\label{Af8}}
\end{figure}

\subsection{Advection of the Pressure Balance Mode}
\label{A4}
The advection of the pressure balance mode can generate unphysical structures, as reported in \citet{kudoh2016}. The initial condition of the test for the left side is $\mathbf{u}_L=(1.0,1.0,0.1,1.0)$, and for the right side is $\mathbf{u}_R=(1.0,1.0,1.0,0.1)$. The simulation end time is $t=2$. This test is conducted in 1D with a spatial grid spacing of $\Delta x = 0.1$. According to \citet{kudoh2016}, the numerical diffusion of non-thermal fluid density leads to an energy loss, hence generating spurious structures. As shown by the black dots in Figure \ref{Af6}, which indicate the total pressure, spurious structures are also found at the discontinuity region due to numerical diffusion. Although we obtain the evolution of the non-thermal fluid pressure by SDE, this structure does not propagate. Additionally, due to the high-order accuracy of the HOW-MHD code, adopting low-order differencing in calculating the gradient of non-thermal fluid pressure generates oscillations at the discontinuity of thermal and non-thermal fluid pressure (see the left panel of Figure \ref{Af6}). This occurs due to a mismatch between the spatial accuracy and the accuracy of the non-thermal fluid pressure gradient. As shown in Figure \ref{Af6}, when implementing high-order differencing, these oscillations are almost suppressed. Hence, we adopt sixth-order differencing as a fiducial scheme.

\subsection{The energy spectrum of accelerated particles by diffusive shock acceleration}
\label{A5}
The strength of solving Parker's equation lies in its ability to obtain the energy spectrum of non-thermal fluid. To demonstrate the proper occurrence of first-order Fermi particle acceleration, we generate a shock with a reflecting boundary. In this test, we generate a parallel shock, with the initial condition given as $\mathbf{u}=(1.0,2.0,0.0,0.0,1.0,0.0,0.0,1.0,P_{\rm{NT}})$. The grid resolution is 256$\times$128 with a domain of [0,1]$\times$[-0.25,0.25]. The right-side boundary is a reflecting boundary, the left side is an open boundary, and the top and bottom boundaries are periodic boundaries for mimicking an infinite shock size. Non-thermal fluid are contained with 1,000 pseudo-particles in each cell. Isotropic spatial diffusion, $\kappa_\perp=\kappa_\parallel$, and the density weighted MHD turbulence diffusion, $\kappa_\parallel=\kappa_0 (p_{\rm{NT}}/p_0)^{4/3}(\rho_g/\rho_1)^{-1}$, are adopted, where $\kappa_0=0.1$ with non-relativistic particles. The density dependence term, $(\rho_g/\rho_1)^{-1}$, where $\rho_1$ is the upstream plasma density, is adopted to quench the energetic particle acoustic instability in the precursor of shock, which is highly modified by non-thermal energetic particles \citep{kang2007}. The initial momentum of pseudo-particles is given as a delta function at $p_0$.
We set the non-thermal fluid pressure of three cases as $P_{\rm{NT},i}=0P_{g,i}$, $0.5P_{g,i}$, and $1P_{g,i}$. The case where $P_{\rm{NT},i}=0P_{g,i}$ does not include feedback from the non-thermal fluid. The simulation end time is $t=0.5$.
To obtain the proper result in diffusive shock acceleration through SDE, the following two conditions shall be satisfied \citep{kong2017}: the first condition is $\kappa_{xx}/v_s>X_{\rm{sh}}$, where $v_s$ and $X_{\rm{sh}}$ are the shock speed and the shock thickness, respectively. The second condition is a small $\Delta t_{\rm{SDE}}$ to capture the transition of the shock. Thus, we adopt the timestep for SDE as $\Delta t_{\rm{SDE}}=10^{-5}(p/p_0)^{-4/3}$, ensuring satisfaction of this relation. The thickness of exponential decrease, $X_{\rm{exp}}$, is determined as $\kappa_2/v_2$, where $\kappa_2$ and $v_2$ are the spatial diffusion coefficient in the downstream and speed of the downstream in the shock rest frame \citep{zhang2000}, respectively. Since this value is a few factors larger than the thickness of the shock, $X_{\rm{exp}}>X_{\rm{sh}}$, pre-compression region  (see the right panel of Figure \ref{Af7}) is generated in the plasma density profile \citep{vink2020}.

As shown in Figure \ref{Af7}, the compression ratio of the shock decreases for the large $P_{\rm{NT},i}$ case. Hence, the energy spectrum of the non-thermal fluid becomes steeper. The dashed line in the left panel of Figure \ref{Af7} is given as $f(p_{\rm{NT}})\propto p_{\rm{NT}}^{-\frac{3\chi}{\chi-1}}$, where $\chi$ is the compression ratio of the shock. It matches very well with the spectral shape of non-thermal fluid. Even though the Mach number of the shock decreases with increasing $P_{\rm{NT,i}}$, the sound speed of the upstream region increases with increasing $P_{\rm{NT,i}}$, given as $c_s=((\gamma_gP_g+\gamma_{\rm{NT}}P_{\rm{NT}})/\rho_g)^{0.5}$, so the shock speed in the upstream rest frame increases. This trend is also consistent when $P_{g,i}$ is 1.5 or 2, but without $P_{\rm{NT,i}}$ (not shown here).

\subsection{Scaling test}
\label{A6}
To estimate the computational efficiency of this method, we perform performance tests focusing on the number of particles in each cell and weak scaling. For these tests, we conduct a reflection shock generation test (section \ref{A5}) in 1D to 3D, with a 2D setup selected for the weak scaling test. In the weak scaling test, the number of particles in each cell is fixed to 8. In the left panel of Figure \ref{Af8}, similar to the MHD-PIC method \citep{sun2023}, the computational cost remains relatively flat when the number of particles is less than $\approx10$, due to constant overhead. Beyond that, the computational cost almost linearly depends on the number of particles in the cell. As shown in the right panel of Figure \ref{Af8}, solving MHD with the SDE method has almost similar efficiency to using only the MHD method.

\bibliography{refs}{}
\bibliographystyle{aasjournal}

\end{document}